\documentclass[preprintnumbers,amsmath,amssymb,aps,floatfix]{revtex4}
\usepackage{txfonts}
\usepackage{graphics}
\usepackage{graphicx}
\usepackage{dcolumn}
\usepackage{bm}
\usepackage{verbatim}
\usepackage{epsfig}
\usepackage{subfigure}
\usepackage{amsmath,amsfonts,amssymb,graphicx,times}

\begin{document}

\title{Degree-layer theory of network topology}
\author{Bin Zhou$^{a,b,\ast}$}
\email{binzhou@mail.ustc.edu.cn}
\author{Bing-Hong Wang$^{a,\ast}$}
\email{bhwang@ustc.edu.cn}
\author{Zhe He$^{a}$}
\affiliation{$^{a}$ Department of Modern Physics, University of Science and Technology of China, Hefei 230026, China}
\affiliation{$^{b}$ Suzhou Insitute of Technology, Jiangsu University of Science and Technology, Zhenjiang 212003, China}

\date{\today}

\begin{abstract}
The network topology can be described by the number of nodes and the interconnections among them. The degree of a node in a network is the number of connections it has to other nodes and the degree distribution is the probability distribution of these degrees over the whole network. Therefore, the degree is very important structural parameter of network topology. However, given the number of nodes and the degree of each node in a network, the topology of the network cannot be determined. Therefore, we propose the degree-layer theory of network topology to describe deeply the network topology. First, we propose the concept of degree-tree with the breadth-first search tree. The degrees of all nodes are layered and have a hierarchical structure. Second,the degree-layer theory is described in detail. Two new concepts are defined in the theory. An index is proposed to quantitatively distinguish the two network topologies. It also can quantitatively measure the stability of network topology built by a model mechanism. One theorem is given and proved, furthermore, and one corollary is derived directly from the theorem. Third, the applications of the degree-layer theory are discussed in the ER random network, WS small world network and BA scale-free network, and the influences of the degree distribution on the stability of network topology are studied in the three networks. In conclusion, the degree-layer theory is helpful for accurately describing the network topology, and provides a new starting point for researching the similarity and isomorphism between two network topologies.
\end{abstract}


\maketitle


\section*{Introduction}

Network science is an emerging, highly interdisciplinary research area that aims to develop theoretical and practical approaches and techniques to increase our understanding of natural and man-made networks. The last decade has witnessed the birth of a new movement of interest and research in the study of complex networks\cite{newman2003structure,boccaletti2006complex,barabasi2009scale,liu2011controllability}. The study of complex networks is pervading all kinds of sciences today, ranging from physical to biological, even to social sciences. Many real complex networks have emerged some common characteristics, such as small world phenomena\cite{watts2003six}, scale-free properties with power-law degree distributions\cite{barabasi2003scale}, and so on. Therefore some important simulation networks with real networks characteristics have been proposed, for instance: ER random network\cite{erd6s1960evolution}, WS small world network\cite{watts1998collective} and BA scale-free network\cite{barabasi1999emergence}, etc. Duncan J. Watts think network science is one of the  most important research subjects in the twenty-first century\cite{watts2007twenty}.

The development of network science must depend on the precise anatomy of network topology. The reason is that the structure always affects function and the behavior of a dynamical system. For instance, the topology of social networks affects the spread of information and disease\cite{newman2002spread}, and the topology of the power grid affects the robustness and stability of power transmission\cite{strogatz2001exploring}. The topology of a network often also plays a crucial role in determining its dynamical features\cite{zhang2006complex}. The apparent ubiquity of complex networks leads to a fascinating set of common problems concerning how the network structure facilitates and constraints the network behaviors. Therefore, it is very important to characterize the structure of complex networks. The research on complex networks begun with the effort of defining concepts and measures to characterize the topology of real networks, such as the degree distributions, degree correlations, excess average degree, average path length, network diameter, clustering coefficient, betweenness, modularity, etc.

Generally speaking, a network topology can be simply defined as a set of interconnected nodes, where a node is a basic element or a fundamental unit with detailed contents depending on the nature of the specific network under consideration. However, network structure is irregular, complex and dynamically evolving in time. With a long tradition, network topology has been studied in graph theory\cite{bollobas1998modern} and discrete mathematics\cite{brandes2005network}. So far, the study of network topology is great difficult and still not enough. The exiting concepts and measures cannot accurately describe the network topology. For instance, the parameters are fixed in the ER random network, WS small world network or BA scale-free network, and a lot of networks can be built and the topologies of them are very different; the self-similarity of complex networks has been researched\cite{song2005self}, but the study of the similarity and isomorphism between two complex networks is very few. Therefore, in order to adapt to the demands of the network science development, we should define new concepts and measures to better characterize the topology of network.

The objective of this article is to propose the degree-layer theory of network topology to describe deeply the network topology. First, The degree-layer theory is described in detail. Two special concepts are defined in the theory. A index is proposed to quantitatively measure the similarity of $n$-layer degrees between two network topologies and quantitatively distinguish the two network topologies. It also can quantitatively measure the stability of network topology built by a model mechanism. One theorem is given and proved. Furthermore, one corollary is derived directly from the theorem. Second, the applications of the degree-layer theory are researched in the ER random network, WS small world network and BA scale-free network. The similarity of one-layer degrees and the similarity of two-layer degrees are discussed in the three networks. In addition, the influences of the degree distribution on the similarity of $n$-layer degrees are discussed in the ER random network, WS small world network and BA scale-free network. Third, conclusions are given. The last part is our acknowledgments, and a number of references are provided at the end of the article.

\section*{Degree-layer theory of network topology}

In a connected network, a node of the network topology can be taken as a root, signed $i$. Starting from the root, a breadth-first search degree tree can be built, signed $T_i$. The $T_i$ is a multi-layer structure and contains all nodes of the network topology, but it does not contain the degree of each node. We make a special provision that the degree of each node is also contained in the breadth-first search tree, and the breadth-first search degree-tree with degrees of all nodes is named degree-tree, signed $D_{i}$. Therefore, the degrees of all nodes are layered with the breadth-first search tree and have a hierarchical structure. The degree trees of all nodes can be composed to a forest, signed $F=\{D_i, i=1,..., M\}$, where $M$ is the total number of all nodes in the network topology. The forest $F$ can characterize the network topology.

Two concepts of the degree-layer theory are defined as follows:
\begin{enumerate}
\item[i)] $n$-layer degree: The previous $n$ layers of the $D_i$ is named n-layer degree of the $D_i$, signed $K_i^n$.

\item[iii)] Identical $n$-layer degree: For $\forall K_i^n, K_j^n$, if $K_i^n$ and $K_j^n$ are isomorphic that $K_i^n \cong K_i^n$, and the degrees of any two nodes which are a one-to-one correspondence in the $K_i^n$ and $K_j^n$ are equal, then we define $K_i^n$ and $K_j^n$ are identical $n$-layer degrees.
\end{enumerate}

For one-layer degree that $n=1$, $K_{i}^{n}=K_i^1$, and $F^n=F^1=\{K_i^1, i=1,..., M\}$. Therefore, one-layer degree $K_i^1$ is the general degree of node $i$ and $F^1$ is a set of all general degrees for all nodes. General degree is a special case in the degree-layer theory. Because $K_{i}^{n}$ is the previous $n$ layers of the $D_i$, the $n$-layer degrees of all degree trees can be composed to a set, signed $F^n=\{K_i^n, i=1,..., M\}$. $F^n$ can characterize the topology of the network, and the larger the $n$ is, the better the characterization of the $F^n$ is. There are two $F_1^n, F_2^n$, which are built from two network topologies $G_1, G_2$, respectively. According to the concept of the identical $n$-layer degree, A index is defined to quantitatively measure the similarity of $n$-layer degrees between $F_1^n$ and $F_2^n$ as follows:
\begin{align}
S^n=\frac{2\alpha^n}{M+N}, \qquad (0 \leq S \leq 1)
\end{align}
where $S^n$ is the similarity of $n$-layer degrees between $F_1^n$ and $F_2^n$, and $\alpha^n$ is the pairs of identical $n$-layer degrees between $F_1^n$ and $F_2^n$. Once a $n$-layer degree is paired, it cannot be paired again. $M, N$ are the total number of all nodes in the two network topologies $G_1, G_2$, respectively. The $S^n$ can also be used to quantitatively distinguish the topologies of the two network $G_1, G_2$, and the larger the $n$ is, the better the distinguishability of $S^n$ is. The schematic illustration of the degree-layer theory is in Fig.1.

One theorem of the degree-layer theory is stated as follows:
\begin{enumerate}
\item[] $ S^{n-1} \geq S^n (n \geq 2)$, for $\forall G_1, G_2$.

Proof.\\
      For $\forall G_1, G_2$, \quad $S^{n-1}$ and $S^n$ can be written as:\\

      \begin{align}
       S^{n-1}=\frac{2\alpha^{n-1}}{M+N}, \qquad S^n=\frac{2\alpha^n}{M+N},
      \end{align}

      where $\alpha^{n-1}$ is the pairs of identical $(n-1)$-layer degrees between $G_1^n$ and $G_2^n$, and $\alpha^n$ is the pairs of identical $n$-layer degrees between $G_1^n$ and $G_2^n$. The two definitions of $n$-layer degree and identical $n$-layer degree imply that:\\

      \begin{align}
       \alpha^{n-1} \geq \alpha^n,
      \end{align}

      then we get\\

      \begin{align}\label{S}
       S^{n-1} \geq S^n.
      \end{align}
\end{enumerate}

One corollary of the theorem is stated as follows:
\begin{enumerate}
\item[] If $S^n =1$, then $S^{n-1}=1(n \geq 2)$, for $\forall G_1, G_2$.

Proof.\\
      If $S^n =1$, we invoke the result of Theorem 1 that\\
      \begin{align}
       S^{n-1} \geq S^n,
      \end{align}
      \begin{align}
       \because \quad 0 \leq S^{n-1} \leq 1, \qquad  0 \leq S^n \leq 1,
      \end{align}
      \begin{align}
       \therefore \quad S^{n-1}=1.
      \end{align}

\end{enumerate}

\section*{The application of the degree-layer theory}

The degree-layer theory of network topology have been described above in detail. In order to show the superiority of the degree-layer theory in describing the network topology, we will study the application of the degree-layer theory in the ER random network, WS small world network and BA scale-free network. In papers\cite{erd6s1960evolution,watts1998collective,barabasi1999emergence}, the three generation mechanisms of the WS small world network, ER random network and BA scale-free network are displayed in detail. The Monte Carlo methods is used in the three generation mechanisms. Therefore, even if the parameters are constant, each mechanism can built a lot of network topologies which are different among them. According to the degree-layer theory, we specially calculate the  $S^1$ and $S^2$ to quantitatively distinguish the different network topologies built by the same model mechanism.

\begin{figure}
\centering
\scalebox{0.4}[0.4]{\includegraphics[bb= 0 0 840 584]{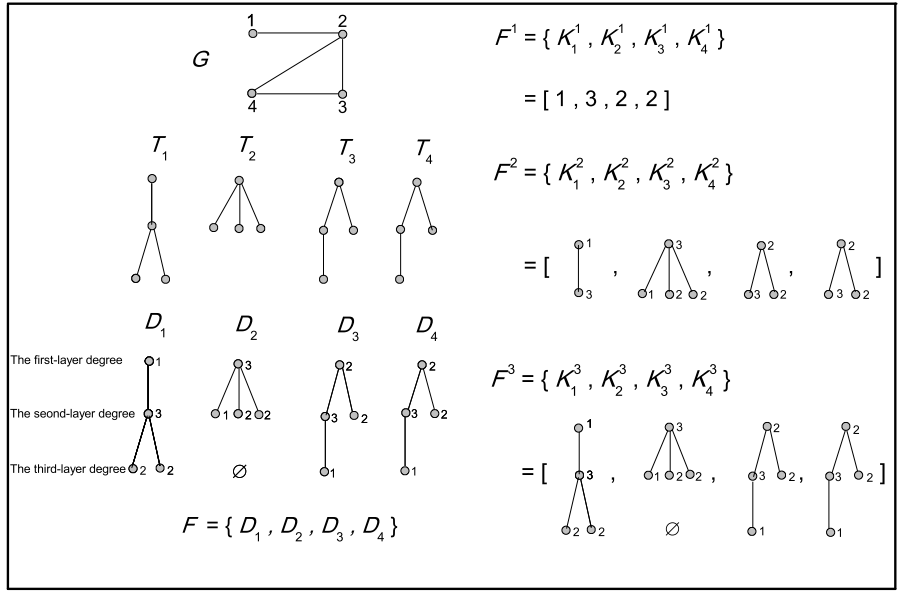}}
\caption{ The schematic illustration of the degree-layer theory. $G$ is a graph with four nodes and four edges. The number which is on the right of each node is the degree of the node in the $D_1, D_2, D_3$ and $D_4$. The $D_2$ is a two-layer degree tree, and the third-layer degree of $D_2$ is defined null, signed $\varnothing$. The $K_2^2, K_3^2, K_3^3$ are correspondingly identical to the $K_2^3, K_4^2, K_4^3$, respectively.}
\end{figure}

\begin{figure}
\centering
\scalebox{0.7}[0.7]{\includegraphics[bb= 20 350 840 584]{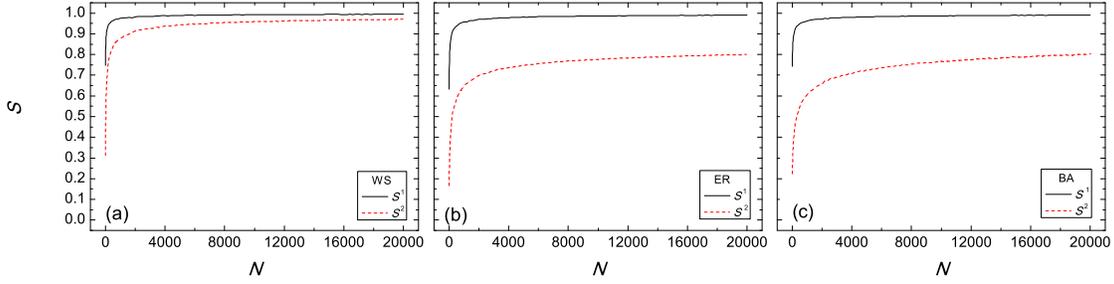}}
\caption{(color online) The comparisons between the $S^1$ and the $S^2$ in the ER random network, WS small world network and BA scale-free network. (a), (b) and (c) represent the corresponding results in the WS small world network, ER random network, BA scale-free network, respectively. In the three sub-figures, $N$ represents the number of nodes and the range of $N$ is from $10$ to $20000$; the black solid line and red dashed line represent the corresponding results of $S^1$ and $S^2$, respectively. In the WS network, the average degree is $2$ and the probability of rewiring each edge at random is $0.5$. In the ER random network, the probability of connecting any two nodes is $2/N$. In the BA scale-free network, starting with a globally coupled network of three nodes, we add a new node with $1$ edge.}
\end{figure}

\begin{figure}
\centering
\scalebox{0.4}[0.4]{\includegraphics[bb= 0 20 841 584]{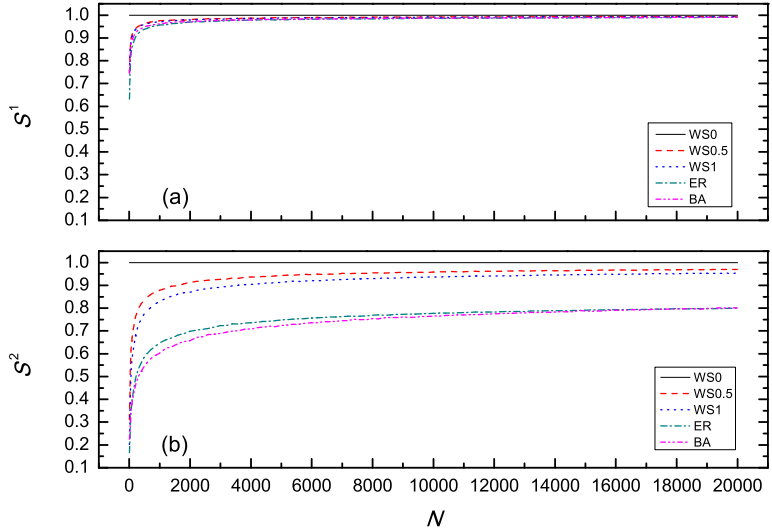}}
\caption{(color online) The comparisons among the $S^1$ and the comparisons among the $S^2$ in the ER random network, WS small world network and BA scale-free network. (a) shows the comparisons among the $S^1$ and (b) shows the comparisons among the $S^2$. The range of $N$ is from $10$ to $20000$. In the two sub-figures, the black solid line represents the results of the WS small network in which the average degree is $2$ and the probability of rewiring each edge at random is $0$, signed WS0; the red dashed line represents the results of the WS small network in which the average degree is $2$ and the probability of rewiring each edge at random is $0.5$, signed WS0.5; the blue dot line represents the results of the WS small network in which the average degree is $2$ and the probability of rewiring each edge at random is $1$, signed WS1; the olive dash-dot line represents the results of the ER random network in which the probability of connection between any two nodes is $2/(N-1)$, signed ER; the magenta dash-dot-dot line represents the results of the BA scale-free network in which starting in a globally coupled network with three nodes, we add a new node with $1$ edge, signed BA}
\end{figure}

\begin{figure}
\centering
\scalebox{0.6}[0.6]{\includegraphics[bb= 0 250 841 534]{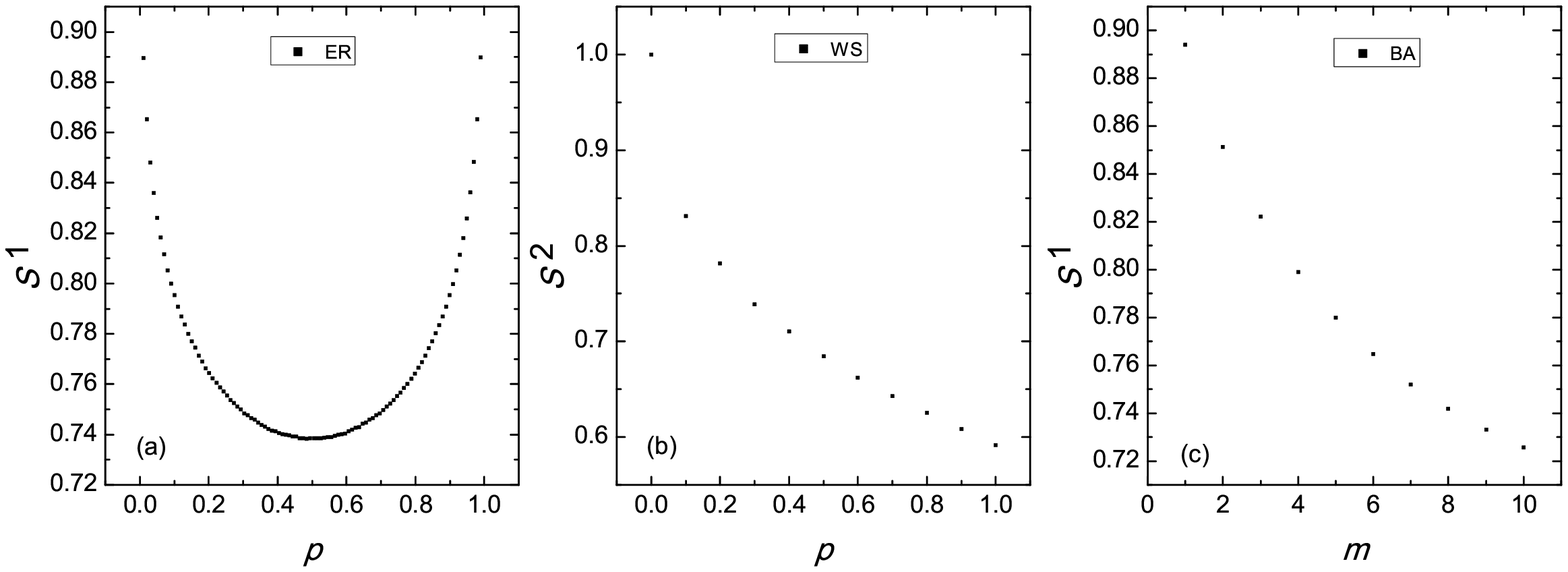}}
\caption{ The influences of the degree distribution on the $S^1$ and $S^2$ in the ER random network, WS small world network and BA scale-free network. (a), (b) and (c) represent the corresponding results of ER random network, WS small world network and BA scale-free network, respectively. In the three sub-figures, the scale of the network is that $N=100$. In the ER random network, $p$ is the probability of the connection between any two nodes, $p \in (0, 1)$. In the WS small world network, $p$ is the probability of rewiring each edge, $p \in [0, 1]$. In the BA scale-free network, starting with a globally coupled network of $m_0$ nodes, at every time step we add a new node with $m$ edges, $m \in [1, 10], m_0=2m+1$.}
\end{figure}

Fig.2 shows the comparisons between the $S^1$ and the $S^2$ in the WS small world network ,ER random network and BA scale-free network. In the three sub-figures, for $\forall N$, $S^1 \geq S^2$ verifies the Theorem 1. $S^1$ more quickly reaches a stable value than $S^2$. When $N > 20$, $S^1 \rightarrow 1$ means the one-layer degrees between the two network topologies are almost identical, and $S^1$ can not distinguish the two network topologies well. However, When $N = 20$, $S^2 < 0.5$ means the the two-layer degrees between the two network topologies are much different, and the $S^2$ can distinguish the two topologies well. When $ N = 20000$, $S^2 \rightarrow 1$, this means that the two-layer degrees between the two network topologies are almost identical, and the $S^2$ can not also distinguish the two network topologies well. In this case, we need to calculate the $S^3$ to better distinguish the two networks topologies. Therefore, in the degree-layer theory, $F^n$ can characterize the topology of a network, and the larger the $n$ is, the better the characterization of $F^n$ is; $S^n$ can quantitatively distinguish two network topologies, and the larger the $n$ is, the better the distinguishability of $S^n$ is.

Fig.3 shows the comparisons among the $S^1$ and the comparisons among the $S^2$ in the ER random network, WS small world network and BA scale-free network. Fig.3(a) shows the comparisons among the $S^1$ and Fig.3(b) shows the comparisons among the $S^2$. In Fig.3(a), when $N >1000$, the four curves of WS0.5, WS1, ER and BA networks approximately tend to WS0 that $S^1=1$. Therefore, $S^1$ can not distinguish each network topology well. In Fig.3(b), the four curves of WS0.5, WS1, ER and BA networks very slowly tend to be a stable value, and $S^2$ can distinguish the each network topology well. We easily find that
\begin{align}
\forall N,\quad  S^2_{WS0}>S^2_{WS0.5}>S^2_{WS1}> S^2_{ER} \approx S^2_{BA}.
\end{align}
The inequalities show that the network topology in the WS0 small world network is the most stable and that the network topology in the ER random network or BA scale-free network or is the most unstable. Therefore, $S^n$ can quantitatively measure the stability of network topology built by a model mechanism, and the larger the $n$, the better the measurement of $S^n$ is. The larger the $S^n$ is, the more stable the network topology built by the model mechanism is. When $N>10000$, the values of $S^2$ in WS0.5, WS1, ER and BA networks tend to be stable, and this shows that the network topology built by each model mechanism has been stable in the two-layer degrees. Therefore, if we want to get a stable network topology built by a model mechanism, we need to let the network scale be large enough. That is to say, the larger the network scale built by a model mechanism is, the more stable the network topology is.

Fig.4 shows the influences of the degree distribution on the $S^1$ and $S^2$ in the ER random network, WS small world network and BA scale-free network. Fig.4(a), (b) and (c) represent the corresponding results of the ER random network, WS small world network and BA scale-free network, respectively. In Fig.4(a), we can find that when $P=0.5$, the $S^1$ is the minimum, implying the network topology is the most unstable. The reasons are the following. The degree distribution of the ER random network follows a binomial distribution. The average degree and the variance of degree distribution in the ER random network can be written as:
\begin{align}
\bar{k}=p(N-1),
\end{align}
\begin{align}
\sigma^2_k=p(1-p)(N-1).
\end{align}
Given the network scale $N$ of the ER random network, when $p=0.5$, the $\sigma^2_k$ is the maximum. It implies that the degree distribution width of the ER network is the largest (see Fig.2(a) in \cite{barabasi1999mean}) and the pairs of the identical one-layer degrees between two network topologies built by the ER model are minimal in the case of $p=0.5$. Therefore, $S^1$ is the minimum in the case of $p=0.5$, and the larger the degree distribution width is, the smaller the $S^1$ is. In Fig.4(b), we can find that the larger the probability of rewiring each edge, the smaller of the $S^2$ is. The reason is that the larger the probability of rewiring each edge is, the larger the degree distribution width of the WS network is(see Fig.2(b) in \cite{barabasi1999mean}). The larger the degree distribution width is, the smaller the $S^2$ also is. In Fig.4(c), we can find the larger the $m$ is, the smaller the $S^1$ is. The reason is that the larger the $m$ is, the larger the degree distribution width of the BA network is(see Fig.4(a) in \cite{barabasi1999mean}). Therefore, the degree distribution plays an important role in the stability of the network topology built by a model mechanism, and the larger the degree distribution width is, the more unstable the network topology is.

\section*{Conclusion}

In this paper, we proposed a degree-layer theory to describe deeply network topology. On the one hand, The degree-layer theory is shown in detail. The degrees of all nodes are layered with the breadth-first search tree. Two concepts are defined in the theory. A index is proposed to quantitatively measure the similarity of $n$-layer degrees between two network topologies. One theorem is given and proved. Furthermore, one corollary is derived directly from the theorem. On the other hand, the applications of the degree-layer theory are researched. First, we discussed the comparison between the similarity of one-layer degrees and similarity of two-layer degrees in the ER random network, WS small world network and BA scale-free network, and the results verify the Theorem 1. $S^2$ can better quantitatively distinguish network topology than $S^1$. The larger the $n$ is, the better the distinguishability of $S^n$ is. Second, we discussed the comparisons among the one-layer degrees and the comparisons among the two-layer degrees in the ER random network, WS small world network and BA scale-free network. The $S^n$ can quantitatively measure the stability of network topology built by a model mechanism, and the larger the $n$, the better the measurement of the $S^n$ is. The larger the $S^n$ is, the more stable the network topology built by the model mechanism is. In addition, the larger the network scale built by a model mechanism is, the more stable the network topology is. Finally, the influences of the degree distribution on the $S^1$ and $S^2$ are discussed in the ER random network, WS small world network and BA scale-free network. The larger the degree distribution width is, the more unstable the network topology is. In conclusion, the degree-layer theory is helpful for describing deeply the network topology, and provides a new starting point for researching the similarity and isomorphism between two network topologies.

\section*{Acknowledgments}

This work is supported by: The National Natural Science Foundation of China (Grant Nos. £º11275186, 91024026). The numerical calculations in this paper have been done on the supercomputing system in the Supercomputing Center of University of Science and Technology of China.

\end{document}